\documentclass[conference]{IEEEtran}
\IEEEoverridecommandlockouts
\usepackage{cite}
\usepackage{amsmath,amssymb,amsfonts}
\usepackage{algorithmic}
\usepackage{graphicx}
\usepackage{textcomp}
\usepackage{xcolor}
\def\BibTeX{{\rm B\kern-.05em{\sc i\kern-.025em b}\kern-.08em
    T\kern-.1667em\lower.7ex\hbox{E}\kern-.125emX}}

\usepackage{hyperref}
\usepackage[noabbrev,capitalise]{cleveref}

\begin{document}

\title{Cherenkov detector with wavelength-shifting fiber readout for muon tomography applications\\}

\author{\IEEEauthorblockN{ Anzori Sh. Georgadze}
\IEEEauthorblockA{\textit{\textsuperscript{1} Institute for Nuclear Research of the National Academy of Sciences of Ukraine, Prospekt Nauky 47, 03680, Kyiv, Ukraine} \\
\textit{\textsuperscript{2} University of Tartu, Ülikooli 18, 50090, Tartu, Estonia }\\
Email: georgadze@kinr.kiev.ua, a.sh.georgadze@gmail.com}
\thanks{Presented at MARESEC 2025 (European Workshop on Maritime Systems Resilience and Security 2025).}
}
\maketitle
\begin{abstract}Cherenkov detectors have been extensively developed and utilized in various scientific fields, including particle physics, astrophysics, and nuclear engineering. These detectors operate based on Cherenkov radiation, which is emitted when a charged particle traverses a dielectric medium at a velocity greater than the phase velocity of light in that medium. In this work, we present the development of a Cherenkov radiation detector designed for a muon tomography system with high spatial resolution, employing wavelength-shifting (WLS) fiber readout. The detector consists of two large-area Cherenkov radiators, each measuring 1~m $\times$ 1~m, with each read out by WLS fibers arranged orthogonally to determine the $x$ and $y$ coordinates of muon hit positions.
The system is modeled using the GEANT4 simulation package, and the achieved position resolution is 1.8 mm$\pm$0.1 (FWHM). 
The detector is inherently insensitive to natural background radiation and exhibits directional sensitivity, which helps prevent image blurring caused by muons scattered from the Earth's surface and entering the detector from the rear. This capability enables applications in non-invasive archaeological and geophysical imaging.
\end{abstract}


\section{Introduction}
Muon radiography and muon tomography are cutting-edge technologies that leverage the unique properties of muons to investigate and visualize the internal structure of large and complex objects.
These methods are increasingly being applied to target as volcanoes, underground cavities, archaeology, glaciers, pyramids, and border security ~\cite{tanaka2007high,lesparre2012density, morishima2017discovery, saracino2017imaging, nishiyama2017first, bonechi,barnes2023cosmic,  yifan2018discrimination, lowZ}. 
Cosmic-Ray Muons are naturally occurring charged particles produced when cosmic rays interact with the Earth's atmosphere. Key advantages of muon tomography non-intrusive inspection in border security is deep penetration of muons, allowing scanning sealed cargo without opening containers. 
Unlike X-ray scanners, which struggle with very dense objects like steel or lead shielding, muon tomography can reveal hidden compartments and smuggled goods.
Muon tomography has been proposed as a tool for border security applications to scan cargo containers and vehicles for hidden contraband or illicit materials~\cite{rapidcargo,antonuccio2017muon,pugliatti2014design,preziosi2020tecnomuse,georgadze2023GEANT4,georgadze2024simulation,georgadze2024,explosives}. 
The European project, "Cosmic beam Tomograph for Identification of Hazardous and Illegal Goods Hidden in Trucks and Sea Containers" (SilentBorder)~\cite{sbwebsite}, focuses on the development and in-situ testing of a high-technology scanner designed for border guards, customs, and law enforcement authorities to inspect shipping containers at border control points.

The use of Cherenkov detectors for muon radiography of volcanoes, buildings, and archaeological sites such as pyramids can be beneficial, as it allows for precise directional determination and prevents image blurring caused by backward-scattered muons.

In muon transmission imaging, a basic muography setup requires at least two position-sensitive planes to track muon trajectories as they pass through an object. The attenuation of muon flux helps reveal variations in density inside the object. In muon scattering tomography (MST), however, the approach is different. Instead of just measuring attenuation, MST analyzes the multiple Coulomb scattering of muons inside the target. To achieve this, the setup requires four tracking detectors - two detectors above the object to measure incoming muon trajectories and two detectors below to track outgoing muon trajectories after interaction. 

\section{Methods}

The Objective of this research is to develop a muon tomography system based on high position resolution detectors. The two types of novel tracking detectors will be developed and constructed. One detector is based on plastic scintillator employing wavelength-shifting (WLS) fiber readout and another is based on the detection of Cherenkov photons produced by cosmic-ray muons in transparent materials, readout using WLS fibers. 

The detector is based on plastic scintillator was preliminary studied in \cite{georgadze2025design}. The detector using Cherenkov photons detection is modelled in this work using GAENT4 toolkit~\cite{AGOSTINELLI}. These detectors are designed as a potential upgrade to the scintillating fiber-based muon tracking detector, which is under development in the SilentBorder project~\cite{sbwebsite}. The key advantage of the Cherenkov detector, designed to detect cosmic ray muons in the range of a few GeV, is its insensitivity to natural radiation background. Additionally, the Cherenkov radiator, made of plastic material, is cheaper, making this option more cost-effective for the construction of a large scale muon tomography station (MTS).

\begin{figure}[t]
\centering
\includegraphics[width=0.5\textwidth]{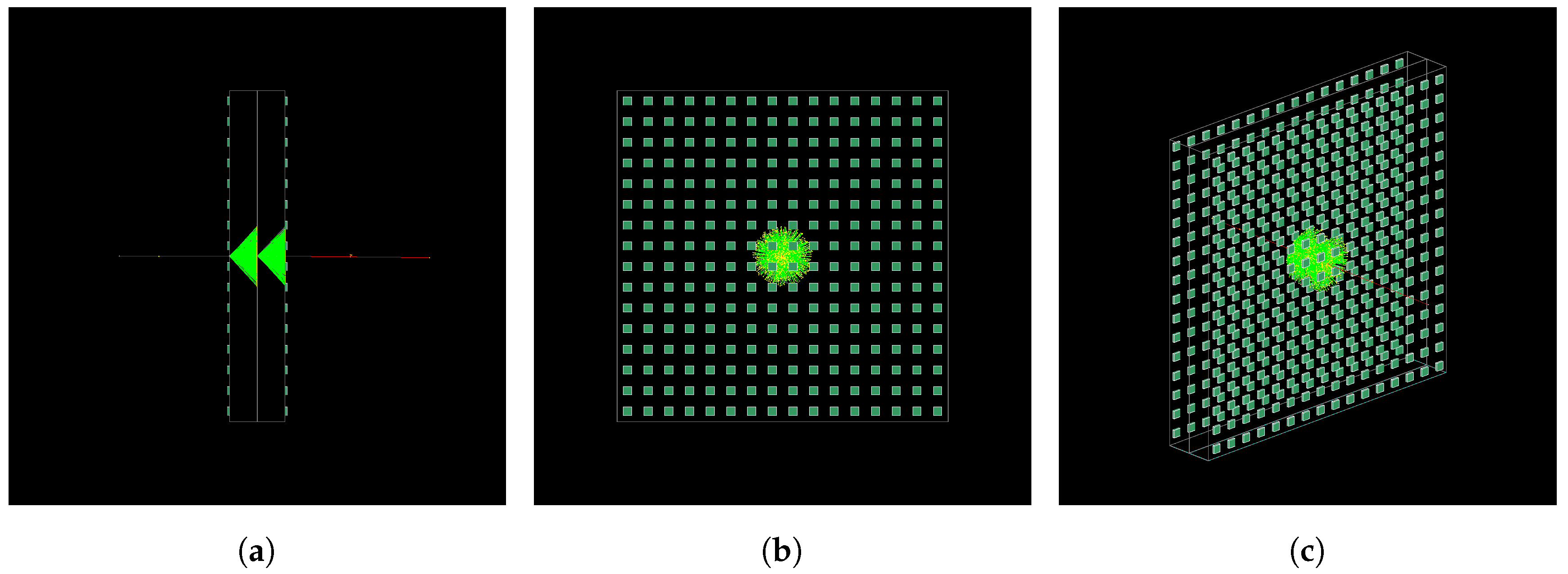}
\caption{Lateral (a), front (b) and perspective visualizations (c) of event of a muon with 10$^5$MeV kinetic energy, simulated in GEANT4. The figures are taken from ~\cite{lo2019feasibility} }
\label{fig:figure1}
\end{figure}

\begin{figure}[t]
\centering
\includegraphics[width=0.156\textwidth]{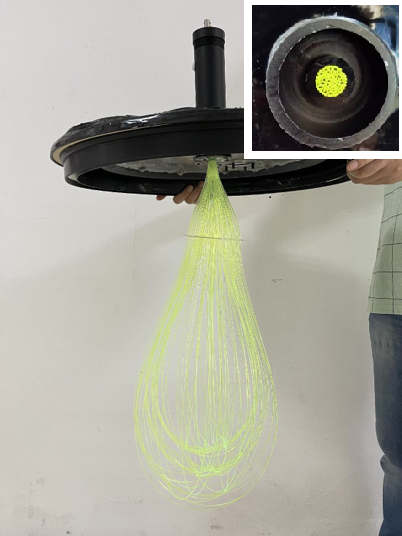}
\includegraphics[width=0.156\textwidth]{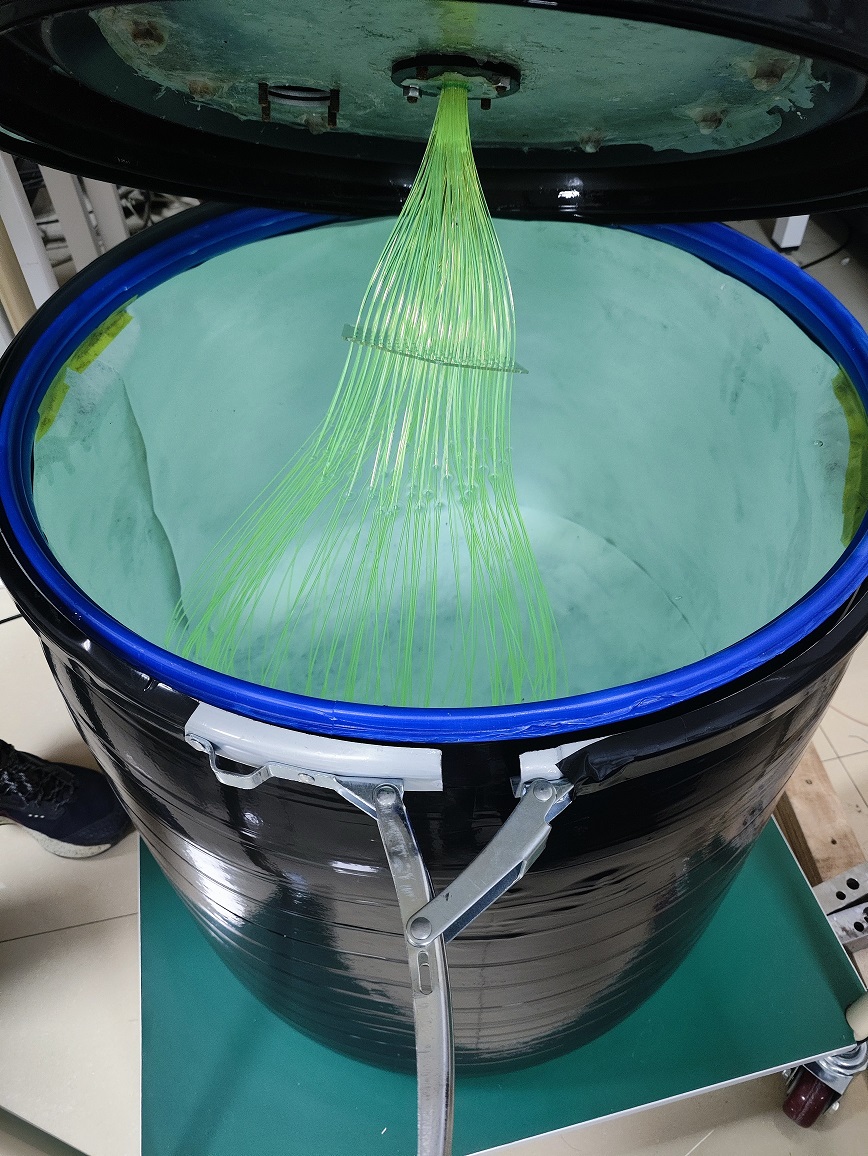}
\includegraphics[width=0.156\textwidth]{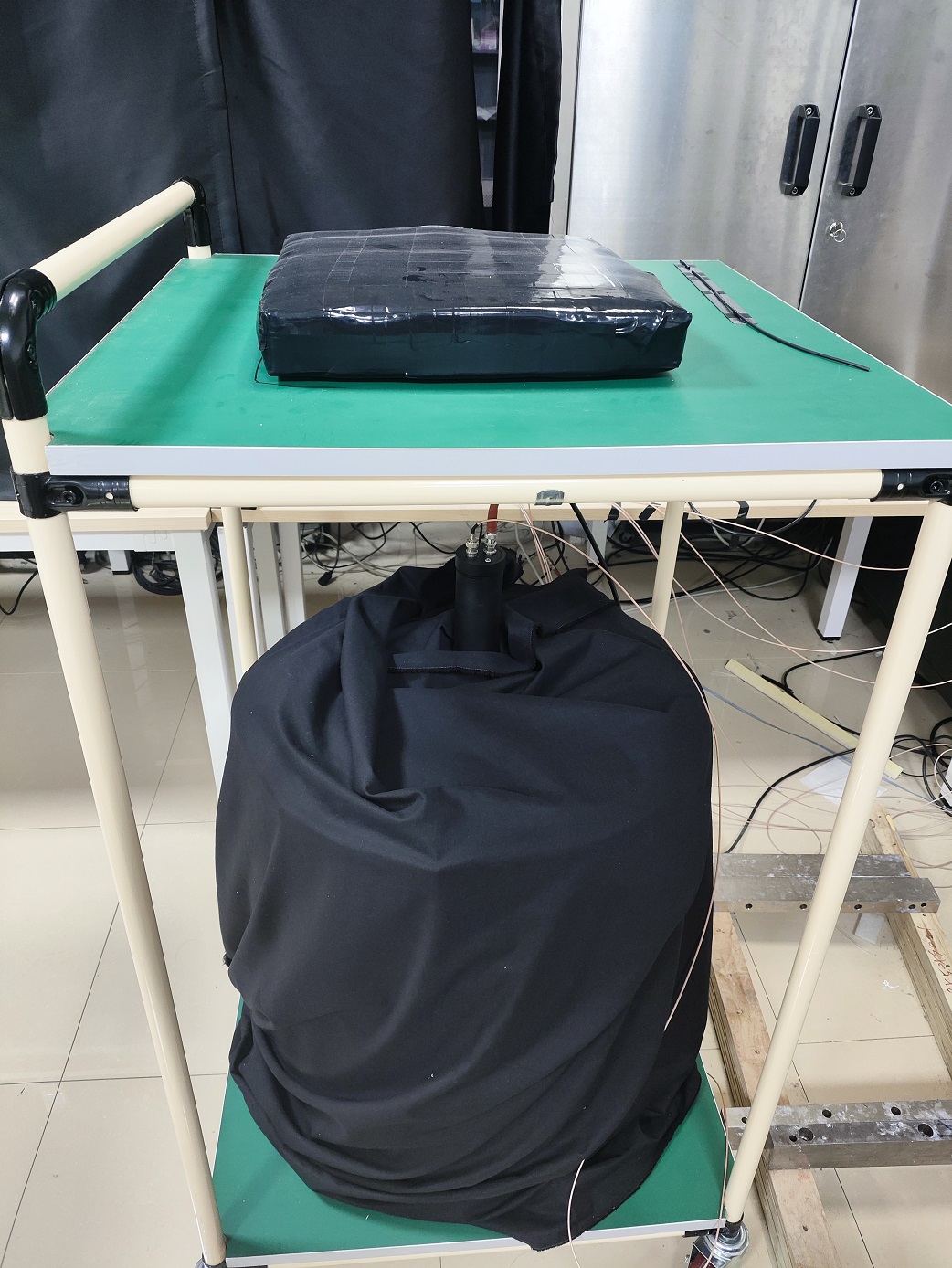}
\caption{(a) Fiber-PMT assembly mounted on the tank lid. The top right corner shows the milled fiber ends. (b) The tank, filled with filtered water; the inner surface was lined with Tyvek 1085D, and the fibers are submerged. (c) Configuration of the experimental setup. Dual trigger detectors are positioned above and below the WCD. The figures are taken from ~\cite{sun2025water}}
\label{fig:figure2}
\end{figure}

\begin{figure}[t]
\centering
\includegraphics[width=0.4\textwidth]{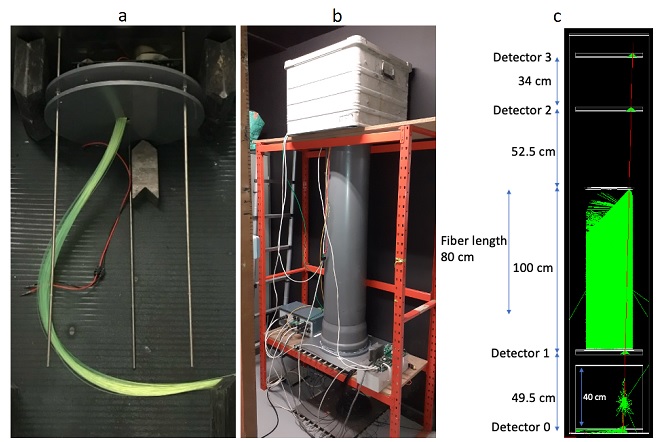}
\caption{(a) The Cherenkov detector light collection system of fibers. The experimental setup (a) and its simulation counter part (c) for testing the DAQ and the Cherenkov response to atmospheric muons. The figures are taken from ~\cite{avgitas2024prototype}}
\label{fig:figure3}
\end{figure}

Cherenkov detectors have been proposed as tracking detectors for muography applications due to their ability to distinguish the direction of incoming muons and thus reject background events from backward-scattered muons~\cite{lo2019feasibility} (see Figure~\ref{fig:figure1}). Another example is using Water Cherenkov detector with fiber enhanced PMT for cosmic ray observation~\cite{sun2025water} (Figure~\ref{fig:figure2}).
Prototype Cherenkov detector for muon tomography applications was also characterized in~\cite{avgitas2024prototype} (Figure~\ref{fig:figure3}). 

Cherenkov detectors rely on the emission of Cherenkov radiation when a charged particle moves through a medium at a speed greater than the phase velocity of light in that medium. 
The number of photons emitted by Cherenkov radiation (per unit length of the particle track) is proportional to:
\begin{align}
\left( 1 - \frac{1}{n(\lambda)^2 \beta^2} \right)
\end{align}
where $n(\lambda)$ is the refractive index of the medium for wavelength  $\lambda$ and $\beta = {v}/{c}$, but only when the expression above is positive (emission is zero for the negative case).
The Cherenkov threshold is usually given in terms of the ratio between the velocity of the particles and the speed of light:
$\beta_{\text{th}} = 1 / {n(\lambda)}$
where $n$ is the refractive index of the medium.  
Since the total energy of the particle is given by:
$E = \gamma m_0 c^2$
and the Lorentz factor is: $\gamma = \left(1 - \beta^2  \right)^{-1/2}$
the Cherenkov threshold energy can be expressed as:
\begin{align}
E_{\text{th}} = m_0 c^2 \frac{1}{\sqrt{1 - \frac{1}{n(\lambda)^2}}}
\end{align}
in which is evident the dependence on particle rest mass $m_0$. The mechanism of Cherenkov effect confines the photons to a cone with its vertex coincident with the point of first light emission. The aperture angle $\theta_c$ of the light cone is related to the particle velocity and to the refractive index of the medium according to the equation
\begin{align}
\cos \theta_c = \frac{1}{\beta_{\text{th}} n(\lambda)}
\end{align}
When a charged particle enters the detector, the emission of Cherenkov photons immediately begins if the particle energy is greater than \( E_{\text{th}} \). The photons produced in the first radiator will be directed toward the WLS-fiber array, where they are absorbed and re-emitted at a longer wavelength isotropically, allowing them to traverse along the WLS fiber due to total internal reflection.  
Then, the muon enters the second radiator, emitting Cherenkov photons that are absorbed in another WLS-fiber array, which is oriented orthogonally to the WLS fibers attached to the first radiator. This configuration enables the determination of the x and y coordinates of the muon’s position.

\section{Results}
\subsection{Description of the Cherenkov detector}
\begin{figure}[t]
\centering
\includegraphics[width=0.5\textwidth]{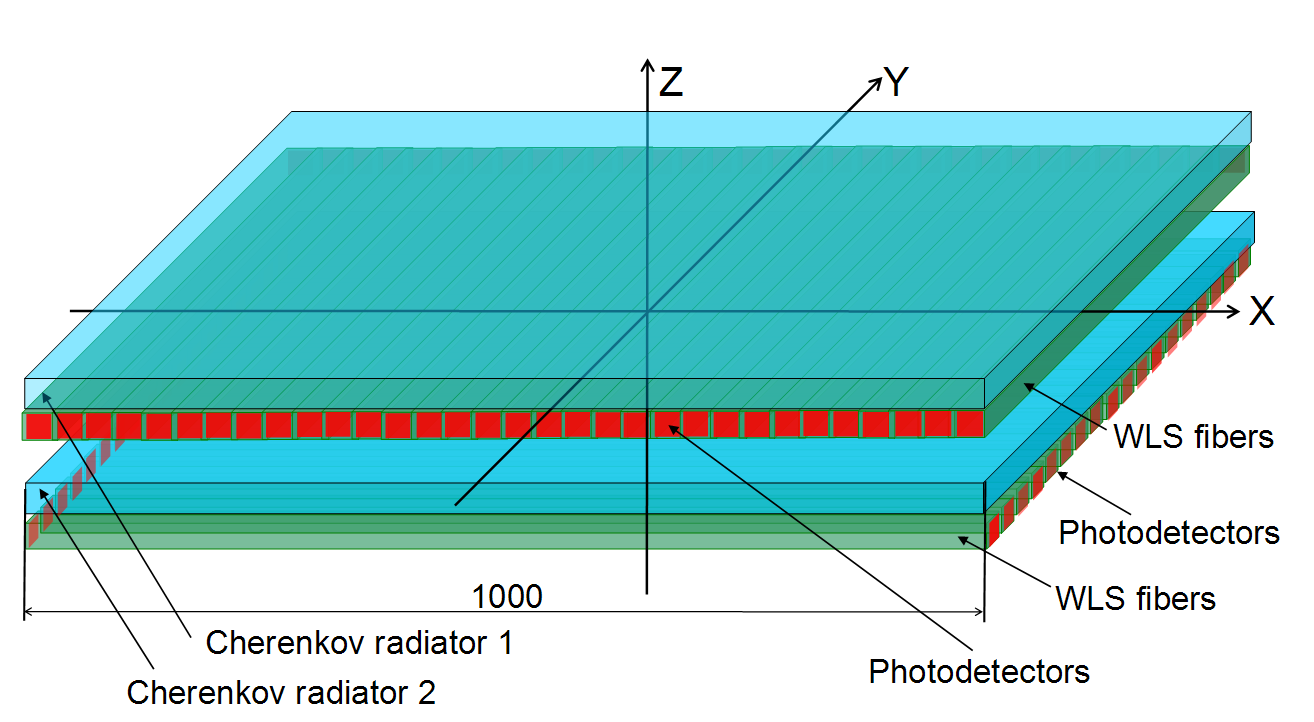}
\caption{Schematic design of a Cherenkov photon detector with WLS-fiber readout (not to scale).}
\label{fig:figure4}
\end{figure}

The Cherenkov detector consists of two radiators, placed one after another. One plate is connected to WLS-fibers in one direction and another plate is connected to WLS-fibers directed in orthogonal direction to allow x and y coordinate determination. 
Fibers are instrumented with silicon photom photomultipliers (SiPM) to detect light. 
The design of the detector is shown in Figure \ref{fig:figure4}. This is a module composed of two radiator of transparent material with size 30$\times$1000$\times$1000 mm$^3$. 
The WLS fibers (2 mm square) attached to 30 mm Plexiglas plates are used to collect photons and shift the UV/blue Cherenkov emission from the Plexiglas and pipe a fraction of the isotropically re-emitted light to SiPMs. 

The material chosen for the radiator was UV-transparent PMMA, commonly known as Plexiglas (C$_5$H$_8$O$_2$, density = 1.19~g/cm$^3$), with a refractive index \textit{n} varying from 1.53 to 1.49 as the photon wavelength increases from 300~nm to 700~nm~\cite{wildner2018analysis}. 
Plexiglas is known for its ability to transmit UV light down to about 300 nm, after which its transmittance decreases significantly ~\cite{nassier2022study, alsaad2021synthesis}. 

\begin{figure}[t]
\centering
\includegraphics[width=0.35\textwidth]{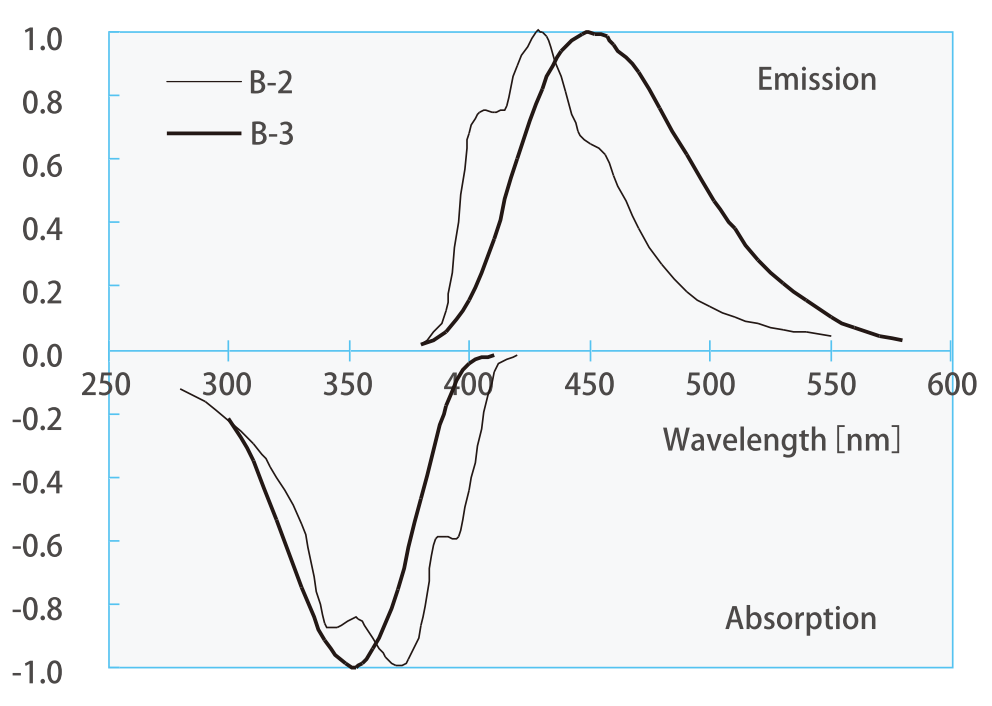}
\caption{The absorption and emission spectra for the WLS fiber B-3(200) used for the light collection of the Cherenkov light~\cite{kuraray} }
\label{fig:figure5} 
\end{figure}

The WLS fibers used were UV-to-blue wavelength-shifting fibers B-3(200) from Kuraray, with an absorption peak at 351 nm and an emission peak at 450 nm. The absorption and emission spectra of B-3(200) are shown in Figure \ref{fig:figure5}. 

\subsection{Simulation of the Cherenkov Detector}
\begin{figure}[t]
\centering
\includegraphics[width=0.45\textwidth]{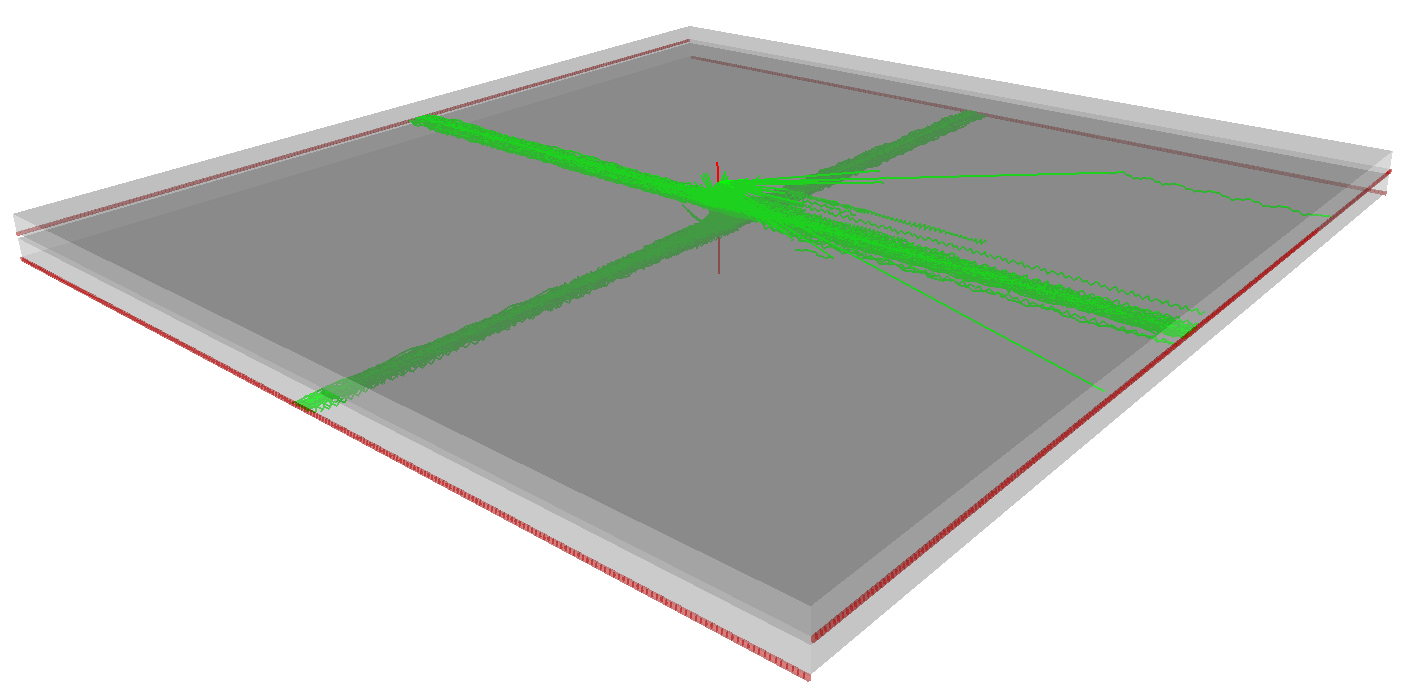}
\caption{Visualization of muon interactions with Cherenkov radiators (30 $\times$ 1000 $\times$ 1000 mm$^3$) simulated in GEANT4, using a configuration that includes a 2~mm square-shaped WLS fiber array and 2 $\times$ 2 mm$^2$ SiPMs (shown in red) for optical readout.}
\label{fig:figure6} 
\end{figure}

\begin{figure}[h]
\centering
\includegraphics[width=0.5\textwidth]{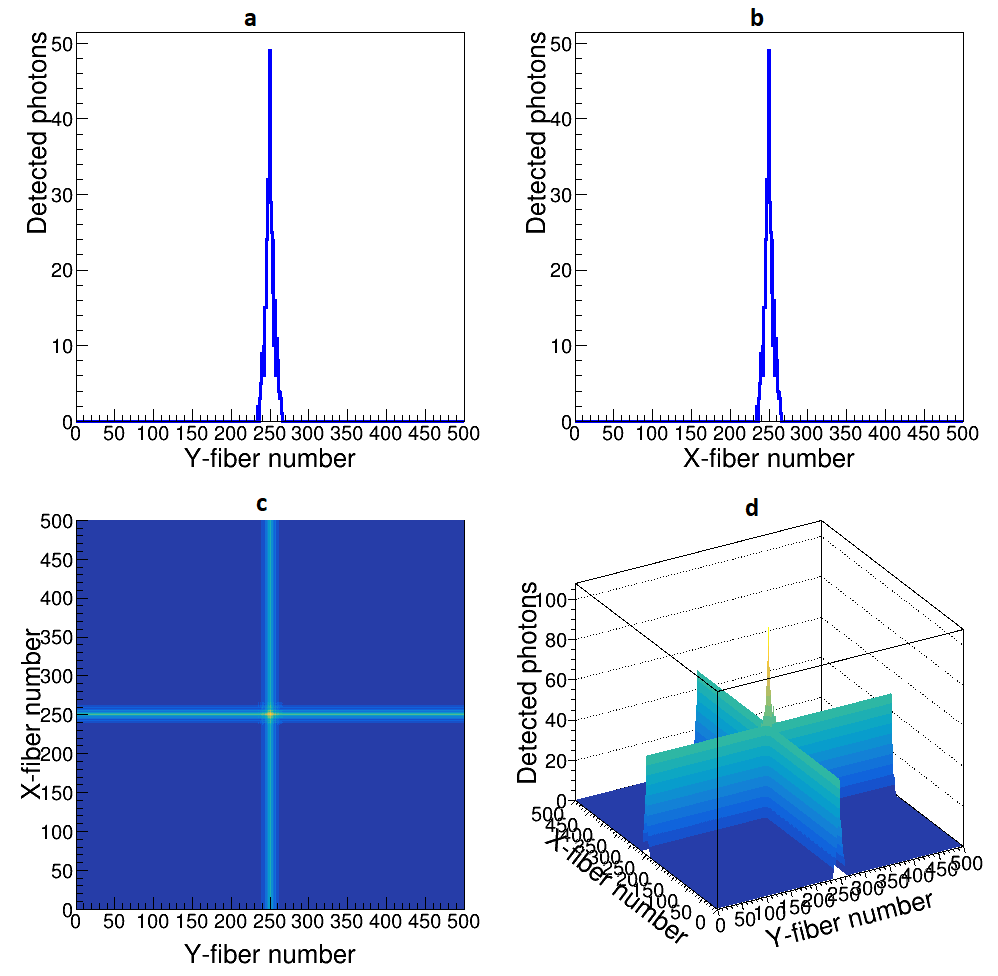}
\caption{(a,b) The light distribution profiles along the $x$ (a) and $y$ (b) directions. (c) The $x–y$ distribution image obtained by combining signals from the $x$ and $y$ WLS fibers. (d) The $x–y$ distribution image in 3D.}
\label{fig:figure7}
\end{figure}

The simulation of optical photons is based on the Monte Carlo code GEANT4. We simulate a muon of 3 GeV energy, passing through the Cherenkov two radiators of the detector. 
Generated Cherenkov photons travel through the plastic radiator plate to the surface attached to WLS fibers where they are absorbed and re-emitted with longer wavelength's, which allow tracking towards SiPMs.
The simulated with GEANT4 Cherenkov photons transport is shown in Figures~\ref{fig:figure6}.
The number of photons detected by each SiPM is recorded event-by-event in ROOT TTree format and analyzed using a C++ code in ROOT environment~\cite{ROOT} . The simulated data represent the distribution of the number of photons detected by the SiPMs connected to both the upper radiator ($\textit{x}$-axis) and lower radiator (\textit{y}-axis) WLS fibers. 
The simulated data represent the distribution of the number of photons detected by the SiPMs connected to both the upper (x axis) and lower (y axis) WLS fibers. The examples
of such distributions are shown in Figure ~\ref{fig:figure7}. 
The ($\textit{x}, \textit{y}$) muon interaction positions are reconstructed using a center of gravity (CoG) algorithm. 

\begin{figure}[h]
\centering
\includegraphics[width=0.5\textwidth]{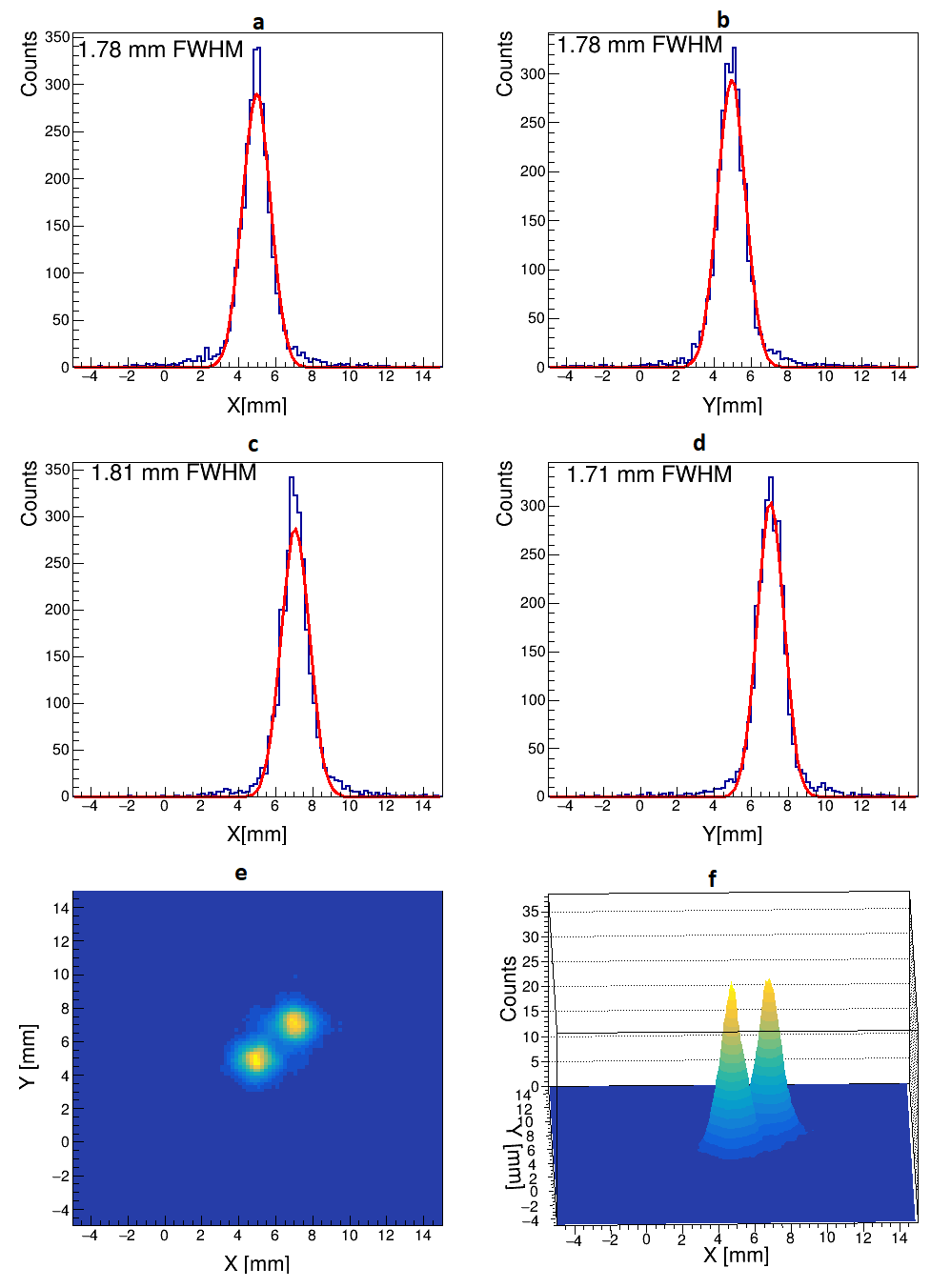}
\caption{Reconstructed muon interaction positions for two muon beam positions impinging on the central detector area, with profiles along the \textit{x}-axis and \textit{y}-axis (a-d), show position resolutions of 1.8 mm (FWHM). The sigma value of the fitted Gaussian waveform can be considered as the position resolution.
Reconstructed muon interaction positions in 2D (e) and 3D (f) representations. Two muon beam coordinates are \textit{x}{$_1$} = 5 mm, \textit{y}{$_1$} = 5 mm and \textit{x}{$_2$} = 7 mm, \textit{y}{$_2$} = 7 mm. }
\label{fig:figure8}
\end{figure}

As shown in Figure~\ref{fig:figure8}, the position resolution achieved is approximately 1.8~mm (FWHM).

Different configurations of the Cherenkov detector were simulated by means of the GEANT4 toolkit, changing the plate thickness and the WLS-fiber cross-section. 
We studied the impact of Cherenkov radiator thickness. A thicker radiator layer results in a larger signal. We found that a 30 mm thickness is acceptable for achieving good spatial resolution.
We also simulated 3 MeV gamma rays impinging on the detector to evaluate its sensitivity to natural background radiation. Simulations show that the Compton electrons produced do not generate a sizable amount of Cherenkov photons compared to those produced by cosmic muons, thus demonstrating the detector's insensitivity to natural background radiation.

\section{Discussion}

An important feature of the proposed detector is that Cherenkov photons are not emitted isotropically, as in the case of scintillators, but are instead emitted at a specific angle that depends on the refractive index \textit{n} of the Cherenkov radiator. Consequently, the detector response is inherently angle-dependent.

Due to the relatively large radiator thickness required to generate a sufficient number of Cherenkov photons (6 cm in this case), the position resolution can be reliably evaluated in a straightforward manner only for muons incident perpendicular to the detector surface. For muons arriving at large inclination angles, position resolution must instead be estimated through full track reconstruction using a multi-module tracking system.

\section{Conclusions}
The detailed geometry of the muon detector, including the layout of the two Plexiglas radiator plates, wavelength-shifting fibers, and SiPMs, is modeled using the GEANT4 toolkit. Simulation results indicate that the proposed detector design, featuring 1 m × 1 m Cherenkov radiators readout with 2 mm$^2$ square-shaped WLS fibers, can achieve a spatial resolution of approximately 1.8 mm$\pm$0.1 (FWHM). 
Reducing the detection surface to 30~$\times~30~\mathrm{mm}^2$ and utilizing optical fibers with a cross-sectional area of 1~$\mathrm{mm}^2$, could be an approach to enhance the spatial resolution of the proposed detector to approximately 1~mm (FWHM).

The results of modeling the Cherenkov detector developed for a muon tomography station, based on photon detection using WLS fibers, show promising results and can also be used for other charged particle tracking applications. 
The detector is inherently insensitive to natural background radiation and exhibits directional sensitivity, which helps prevent image blurring caused by muons scattered from the Earth's surface and entering the detector from the rear side.
The detection efficiency of Cherenkov detectors depends strongly on the incident angle of incoming muons, making their field of view inherently narrow. Considering this feature and their directional sensitivity, Cherenkov detectors are especially suitable for imaging large objects such as volcanoes and pyramids. 
Moreover, their relatively compact design and cost-effectiveness make Cherenkov detectors promising candidates for mobile or modular muography systems. Therefore, they may play an important role in the next generation of non-invasive archaeological and geophysical imaging technologies.

\section{Acknowledgments}
The author acknowledge partial funding from the the EU Horizon 2020 Research and Innovation Programme under grant agreement no. 101021812 ("SilentBorder").

\bibliographystyle{IEEEtranDOI}
\bibliography{biblio.bib}
\end{document}